\documentclass[twocolumn]{aastex631}

\usepackage{graphicx,amsmath}
\usepackage{amssymb}
\usepackage{nccmath}
\usepackage{upgreek}
\usepackage{subfigure}
\usepackage{color}
\usepackage{bm}
\usepackage{soul}
\usepackage{CJK}
\usepackage{ulem}
\usepackage{pifont}
\hypersetup{linkcolor=blue, citecolor=blue}

\setcitestyle{notesep={ }}
\newcommand{\Msol}{\hbox{\thinspace $M_{\odot}$}}

\newcommand{\htwo}{H$_2$}
\newcommand{\hi}{H\thinspace{\sc i}}
\newcommand{\htwohi}{$M_{\rm H_2}/M_{\rm HI}$}

\newcommand{\ha}{H$\alpha$}

\shorttitle{The critical role of dark matter halos in driving star formation}
\shortauthors{Dou et al.}

\begin{document}

\title{{\bf \Large The critical role of dark matter halos in driving star formation}}

\correspondingauthor{Jing Dou, Yingjie Peng, Qiusheng Gu}
\email{doujing@nju.edu.cn, yjpeng@pku.edu.cn, qsgu@nju.edu.cn}
\vspace{-1cm}

\author[0000-0002-6961-6378]{Jing Dou*}
\affiliation{School of Astronomy and Space Science, Nanjing University, Nanjing 210093, China}
\affiliation{Key Laboratory of Modern Astronomy and Astrophysics (Nanjing University), Ministry of Education, Nanjing 210093, China}

\author[0000-0003-0939-9671]{Yingjie Peng*}
\affiliation{Department of Astronomy, School of Physics, Peking University, 5 Yiheyuan Road, Beijing 100871, China}
\affiliation{Kavli Institute for Astronomy and Astrophysics, Peking University, 5 Yiheyuan Road, Beijing 100871, China}

\author[0000-0002-3890-3729]{Qiusheng Gu*}
\affiliation{School of Astronomy and Space Science, Nanjing University, Nanjing 210093, China}
\affiliation{Key Laboratory of Modern Astronomy and Astrophysics (Nanjing University), Ministry of Education, Nanjing 210093, China}

\author[0000-0001-6947-5846]{Luis C. Ho}
\affiliation{Kavli Institute for Astronomy and Astrophysics, Peking University, 5 Yiheyuan Road, Beijing 100871, China}
\affiliation{Department of Astronomy, School of Physics, Peking University, 5 Yiheyuan Road, Beijing 100871, China}

\author[0000-0002-7093-7355]{Alvio Renzini}
\affiliation{INAF - Osservatorio Astronomico di Padova, Vicolo dell'Osservatorio 5, I-35122 Padova, Italy}

\author[0000-0002-8614-6275]{Yong Shi}
\affiliation{School of Astronomy and Space Science, Nanjing University, Nanjing 210093, China}
\affiliation{Key Laboratory of Modern Astronomy and Astrophysics (Nanjing University), Ministry of Education, Nanjing 210093, China}

\author[0000-0002-3331-9590]{Emanuele Daddi}
\affiliation{AIM, CEA, CNRS, Universit\'{e} Paris-Saclay, Universit\'{e} Paris Diderot, Sorbonne Paris Cit\'{e}, F-91191 Gif-sur-Yvette, France}

\author[0009-0001-1564-3944]{Dingyi Zhao}
\affiliation{Department of Astronomy, School of Physics, Peking University, 5 Yiheyuan Road, Beijing 100871, China}
\affiliation{Kavli Institute for Astronomy and Astrophysics, Peking University, 5 Yiheyuan Road, Beijing 100871, China}

\author[0000-0001-6469-1582]{Chengpeng Zhang}
\affiliation{Korea Astronomy and Space Science Institute, Daedeok-daero 776, Yuseong-gu, Daejeon 34055, Republic of Korea}

\affiliation{Department of Astronomy, Yonsei University, Yonsei-ro 50, Seodaemun-gu, Seoul 03722, Republic of Korea}

\author[0000-0002-0182-1973]{Zeyu Gao}
\affiliation{Department of Astronomy, School of Physics, Peking University, 5 Yiheyuan Road, Beijing 100871, China}
\affiliation{Kavli Institute for Astronomy and Astrophysics, Peking University, 5 Yiheyuan Road, Beijing 100871, China}

\author[0000-0003-3010-7661]{Di Li}
\affiliation{Department of Astronomy, Tsinghua University, Beijing 100084, China}
\affiliation{National Astronomical Observatories, Chinese Academy of Sciences, Beijing 100101, China}

\author[0009-0000-7307-6362]{Cheqiu Lyu}
\affiliation{Department of Astronomy, School of Physics, Peking University, 5 Yiheyuan Road, Beijing 100871, China}
\affiliation{Kavli Institute for Astronomy and Astrophysics, Peking University, 5 Yiheyuan Road, Beijing 100871, China}

\affiliation{CAS Key Laboratory for Research in Galaxies and Cosmology, Department of Astronomy, University of Science and Technology of China, Hefei, Anhui 230026, China}

\affiliation{School of Astronomy and Space Science, University of Science and Technology of China, Hefei 230026, China}

\author[0000-0002-4803-2381]{Filippo Mannucci}
\affiliation{Istituto Nazionale di Astrofisica, Osservatorio Astrofisico di Arcetri, Largo Enrico Fermi 5, I-50125 Firenze, Italy}

\author[0000-0002-4985-3819]{Roberto Maiolino}
\affiliation{Cavendish Laboratory, University of Cambridge, 19 J. J. Thomson Avenue, Cambridge CB3 0HE, UK}
\affiliation{Kavli Institute for Cosmology, University of Cambridge, Madingley Road, Cambridge CB3 0HA, UK}
\affiliation{Department of Physics and Astronomy, University College London, Gower Street, London WC1E 6BT, UK}

\author[0000-0002-2504-2421]{Tao Wang}
\affiliation{School of Astronomy and Space Science, Nanjing University, Nanjing 210093, China}
\affiliation{Key Laboratory of Modern Astronomy and Astrophysics (Nanjing University), Ministry of Education, Nanjing 210093, China}

\author[0000-0003-3564-6437]{Feng Yuan}
\affiliation{Center for Astronomy and Astrophysics and Department of Physics, Fudan University, Shanghai 200438, People’s Republic of China}

\begin{abstract}

\noindent Understanding the physical mechanisms that drive star formation is crucial for advancing our knowledge of galaxy evolution. We explore the interrelationships between key galaxy properties associated with star formation, with a particular focus on the impact of dark matter halos. Given the sensitivity of atomic hydrogen (\hi) to external processes, we concentrate exclusively on central spiral galaxies. We find that the molecular-to-atomic gas mass ratio ($M_{\rm H_2}/M_{\rm HI}$) strongly depends on stellar mass and specific star formation rate (sSFR). In the star formation efficiency (SFE)–sSFR plane, most galaxies fall below the \htwo\ fundamental formation relation (FFR), with SFE$_{\rm HI}$ being consistently lower than SFE$_{\rm H_2}$. Using the improved halo masses derived by \citet{2025ApJ...979...42Z}, for star-forming galaxies, both SFE$_{\rm HI}$ and $M_{\rm H_2}/M_{\rm HI}$ increase rapidly and monotonically with halo mass, indicating a higher efficiency in converting \hi\ to \htwo\ in more massive halos. This trend ultimately leads to the unsustainable state where SFE$_{\rm HI}$ exceeds SFE$_{\rm H_2}$ at halo mass around $10^{12} \Msol$. For halos with masses exceeding $10^{12} \Msol$, galaxies predominantly experience quenching. We propose a plausible evolutionary scenario in which the growth of halo mass regulates the conversion of \hi\ to \htwo, star formation, and the eventual quenching of galaxies. The disk size, primarily regulated by the mass, spin and concentration of the dark matter halo, also significantly influences \hi\ to \htwo\ conversion and star formation. These findings underscore the critical role of dark matter halos as a global regulator of galaxy-wide star formation, a key factor that has been largely underappreciated in previous studies.

\end{abstract}

\vspace{-1.5cm}
\section{Introduction} \label{sec:intro} 

Understanding the physical mechanisms that regulate star formation is essential for deepening our comprehension of galaxy evolution. In the $\Lambda$ cold dark matter ($\Lambda$CDM) framework, galaxies reside within dark matter (DM) halos, which provide the gravitational potential wells for gas to cool and condense, ultimately leading to the formation of stars \citep[e.g.,][]{1970A&A.....5...84Z,1977MNRAS.179..541R,1977ApJ...211..638S,1978MNRAS.183..341W,1996ApJ...462..563N,2005Natur.435..629S,2018ARA&A..56..435W}. While the general processes of galaxy formation and the hierarchical buildup of DM halos are well-established, the intricate interplay between DM halos and the baryonic processes that drive star formation in galaxies remains a pivotal area of research in astrophysics.

Central to the process of star formation is the cold interstellar medium (ISM), primarily composed of atomic hydrogen (\hi) and molecular hydrogen (\htwo). These two phases of gas serve distinct roles: \hi\ provides a reservoir of raw material that can eventually contribute to star formation and is typically found in the outer, less dense regions of galaxies \citep[e.g.,][]{2002ApJ...569..157W,2008Leroy,2008AJ....136.2846B,2010AJ....140.1194B}. In contrast, \htwo\ is necessary for star formation and dominates in the denser, inner regions where molecular clouds form \citep[e.g.,][]{1959ApJ...129..243S,1998ApJ...498..541K,2008AJ....136.2846B,2010MNRAS.403..683C,2018MNRAS.476..875C,2012ARA&A..50..531K,2016MNRAS.462.1749S,2022ARA&A..60..319S}. Numerous studies have shown that \hi\ can collapse and convert into molecular gas under suitable conditions, subsequently fueling star formation \citep[e.g.,][]{2006Blitz,2009ApJ...699..850K}. This process is critical, as the presence of \htwo\ is a prerequisite for the formation of stars. Understanding the molecular-to-atomic gas mass ratio ($M_{\rm H_2}/M_{\rm HI}$) is therefore crucial, as it provides insight into the efficiency of this conversion process and its implications for galaxy evolution.

The $M_{\rm H_2}/M_{\rm HI}$ is influenced by various factors, including mid-plane hydrostatic gas pressure \citep[e.g.,][]{2004ApJ...612L..29B,2006Blitz,2008Leroy,2009ApJ...697...55G,2011ApJ...728...88G,2012ApJ...759....9K,2018ApJS..238...33D,2020MNRAS.497..146D}, ultraviolet radiation fields \citep[e.g.,][]{1993ApJ...411..170E}, and the presence of dust \citep[e.g.,][]{2009ApJ...699..850K}. Mid-plane pressure is particularly important, as higher mid-plane pressures often correlate with denser gas regions, facilitating the formation of molecular hydrogen on dust grains under higher pressure and density conditions. Additionally, mid-plane pressure can affect the level of turbulence and the magnetic field strength within the ISM \citep[e.g.,][]{2003ApJ...590L...1K}. Turbulence can promote the mixing and compression of gas, aiding in the formation of molecular clouds \citep[e.g.,][]{2005ApJ...630..250K}, while magnetic fields can provide additional support against gravitational collapse, affecting the balance between \hi\ and \htwo~ \citep[e.g.,][]{1976ApJ...207..141M}.

Despite extensive observational and theoretical work, the mechanisms governing the \hi-to-\htwo\ transition, particularly in relation to DM halo properties, remain insufficiently understood. DM halos play a significant role in shaping the baryonic content of galaxies, as larger halos can accrete more gas and retain gas more efficiently, leading a more substantial \hi\ reservoir and higher central gas densities. Various observations and simulations have established a tight correlation between stellar mass and halo mass, known as the stellar mass-halo mass (SMHM) relation \citep[e.g.,][]{2003MNRAS.339.1057Y,2010ApJ...710..903M,2012ApJ...744..159L,2018AstL...44....8K,2019MNRAS.488.3143B}. This relationship suggests that the growth and assembly of DM halos significantly influence the star formation processes of galaxies. Recent studies have also focused on the relationship between \hi\ content and halo mass \citep[e.g.,][]{2017MNRAS.469.2323P,2018ApJ...866..135V,2019MNRAS.483.4922B,2020MNRAS.498...44C}. However, the specific role of DM halos in regulating star formation remains less well understood.

To accurately assess the role of DM halos in regulating the \hi-to-\htwo\ conversion (i.e., $M_{\rm H_2}/M_{\rm HI}$) and star formation efficiency (SFE, defined as SFR/$M_{\rm gas}$), it is essential to exclude external environmental factors, as \hi\ gas is particularly sensitive to external influences such as ram pressure stripping and galaxy interactions \citep[e.g.,][]{1984Haynes,2001Solanes,2002ApJ...569..157W,2005Gavazzi,2008AJ....136.2846B,2009AJ....138.1741C,2017A&A...605A..18C}. Therefore, we specifically select central spiral galaxies as in \citet{2024ApJ...973L..23D}, as also shown in Section \ref{sec:sample} in detail. This choice minimizes environmental effects, which are more pronounced in satellite galaxies \citep[e.g.,][]{2021PASA...38...35C}. Additionally, galaxies with different morphologies may follow very different evolutionary paths, making it essential to focus on a specific type for clearer insights. For instance, many elliptical galaxies have already quenched at higher redshifts \citep[e.g.,][]{2005ApJ...626..680D,2012ApJ...755...26O}, and including them would introduce significant progenitor bias. Also, ellipticals in the local universe are predominantly transformed from disk galaxies through mergers rather than internal secular evolution \citep[e.g.,][]{1992ApJ...393..484B,2005A&A...437...69B}. For irregulars and S0 galaxies, the formation mechanisms remain unclear and may involve recent mergers or strong interactions \citep[e.g.,][]{2011MNRAS.415.1783B,2017A&A...604A.105T,2018ApJ...862..100G}. Hence, our study focuses exclusively on central spiral galaxies to study the connections between gas content, star formation, and DM halo properties, as this sample serves as the most appropriate laboratory to study the transformation from \hi\ to \htwo\ and its subsequent role in star formation. Specifically, we analyze how the $M_{\rm H_2}/M_{\rm HI}$ and star formation efficiency in both the atomic (SFE$_{\rm HI}$, defined as SFR/M$_{\rm HI}$) and molecular (SFE$_{\rm H_2}$, defined as SFR/M$_{\rm H_2}$) gas phases vary with stellar mass, specific star formation rate (sSFR, defined as SFR/$M_*$), and DM halo mass. We aim to elucidate the role that DM halos play in regulating the conversion of \hi\ to \htwo\ and galaxy-wide star formation.

Throughout this work, we assume the following cosmological parameters: $\Omega_m=0.3, \ \Omega_\Lambda=0.7, \ H_0=70\,\rm {km\,s^{-1} Mpc^{-1}}$.

\section{Sample} \label{sec:sample}

\subsection{The ALFALFA-SDSS matched sample} 

The \hi\ sample used in this work is from Arecibo Legacy Fast ALFA (ALFALFA) survey \citep{2011AJ....142..170H,2018ApJ...861...49H}, which provides a comprehensive view of the \hi\ gas content in the nearby universe out to $z \sim$ 0.06. It surveyed approximately 7000 deg$^2$ and contains over 30,000 extragalactic \hi\ detections. ALFALFA uses a drift-scan technique and it is highly efficient for wide-area surveys. In our analysis, the \hi\ detections sample includes not only high signal-to-noise ratio ($>$ 6.5) \hi\ detections (code 1), but also those matched with optical counterparts of comparable redshift (code 2) although have lower signal-to-noise ratio.

To acquire additional optical properties, the ALFALFA sample was then cross-matched with the parent Sloan Digital Sky Survey Data Release Seven (SDSS DR7) \citep{Abazajian:2009ef} sample. The SDSS sample was initially obtained from the SDSS CasJobs site and constructed following the methodology detailed in \citet{Peng:2010gn,2012ApJ...757....4P}. The sample included galaxies with clean photometry, and Petrosian r magnitudes in the range of 10.0 to 18.0 after correction for Milky-Way galactic extinction. Once duplicates were removed, the parent photometric sample contains 1,579,314 objects, among which 72,697 have reliable spectroscopic redshift measurements within the redshift range 0.02 $< z <$ 0.05. This narrow redshift interval was chosen to match the depth of the ALFALFA survey, ensuring consistency in the analysis. The ALFALFA \hi\ detections are cross-matched with the parent SDSS sample using the following criteria. Firstly, the spatial separation between the most probable optical counterpart of each \hi\ source and SDSS galaxy is less than 5$''$. Also, the velocity difference between the \hi\ detection and SDSS galaxy is less than 300 km/s. Besides, since the angular resolution in ALFALFA is $\sim$ 3.5 arcmin, the measured \hi\ spectra may be contaminated by close companions. We exclude galaxies that have multiple SDSS counterparts within the beam radius and within a velocity difference of three times the \hi\ line width ($W_{50}$). These \hi\ detections with clean SDSS counterparts is referred to as the ALFALFA-SDSS sample.
The ALFALFA-SDSS matched sample contains 9571 reliable \hi\ detections

SDSS and ALFALFA are both flux-limited samples, which produce strong selection bias towards massive, \hi-rich, star-forming galaxies even within the narrow redshift range ($z =$ 0.02 - 0.05). We have performed careful incompleteness corrections to the SDSS and ALFLAFA sample independently and combined them together to correct the ALFALFA-SDSS matched sample, as shown in detail in \citet{2019ApJ...884L..52Z} and \citet{2024ApJ...973L..23D}. In brief, each galaxy is weighted by the value of $V_{\rm total}$/$V_{\rm max}$ to account for the volume incompleteness within the given redshift range, where $V_{\rm total}$ is the total comoving volume that the sample spans, and $V_{\rm max}$ is the maximum observable comoving volume for each galaxy. As illustrated in the lower panels of Figure 1 in \citet{2024ApJ...973L..23D}, the \hi\ detection ratio for central spiral galaxies in the ALFALFA is generally low when selection effects are not accounted for. After applying these corrections, the \hi\ detection ratio increases significantly, becoming comparable to that of the xGASS sample, a deeper \hi\ survey in the local universe (albeit with a much smaller sample size).

\subsection{Selection criteria and properties of ALFALFA-SDSS central spiral galaxies}

The ALFALFA-SDSS galaxies are categorized into central and satellite galaxies based on the SDSS DR7 group catalogue from \citet{Yang:2007}. The central galaxies are defined as the most massive and most luminous ones in the $r$-band within a given group, including both ``centrals with satellites" and ``isolated centrals/singletons".

We identify spiral galaxies based on their visual morphology from the Galaxy Zoo project \citep{2011MNRAS.410..166L}, where a vast number of volunteers evaluate SDSS images and categorize galaxy types. A clean sample is constructed by ensuring that a minimum of 80\% of the adjusted votes align with a single category. Each galaxy is then labeled with a morphology flag (“spiral,” “elliptical,” or “uncertain”) after undergoing a careful debiasing process. Notably, in the Galaxy Zoo project, “spiral” refers to both disky galaxies with prominent spiral arms and those without distinct spiral structures, and we categorize them all as “spirals”. Mergers are excluded from our sample if the vote fractions of a merger are greater than 0.3, as galaxy mergers can have a complicated effect on the star formation of galaxies. As shown in detail in \citet{2024ApJ...973L..23D}, visually-classified “spiral” serves as the most effective parameter for distinguishing \hi-rich galaxies and \hi-poor galaxies. And also, central spiral galaxies defined by visual morphology in the Galaxy Zoo project exhibit a very high \hi\ detection fraction, which is crucial for obtaining unbiased intrinsic \hi\ scaling relations, as shown in \citet{2024ApJ...973L..23D}. 

In total, there are 4470 central spiral galaxies with reliable \hi\ detections in the ALFALFA-SDSS matched sample. The stellar mass ($M_*$) of each galaxy is determined using the $k-correction$ program \citep{2007AJ....133..734B} with \citet{2003MNRAS.344.1000B} stellar population synthesis model. The star formation rates (SFRs) are taken from the value-added MPA-JHU catalog \citep{2004MNRAS.351.1151B}, which are derived from the \ha\ emission line luminosities. These luminosities are corrected for extinction using the H$\alpha$/H$\beta$ ratio. To correct for the aperture effects, the SFRs outside the SDSS 3$''$ fiber were obtained by performing the spectral energy distribution fitting to the $ugriz$ photometry outside the fiber, using the models and methods described in \citet{2007ApJS..173..267S}. For AGN and composite galaxies, the central nuclear activities can contaminate the \ha\ emission. Their SFRs are estimated based on the strength of 4000\textup{~\AA} break, calibrated with H$\alpha$ for non-AGN, pure star-forming galaxies (detailed in \citealt{2004MNRAS.351.1151B}). These SFRs are computed for a Kroupa initial mass function (IMF) and we convert them to a Chabrier IMF using log SFR (Chabrier) = log SFR (Kroupa) - 0.04. We also conducted the same analysis using SFRs derived from the SED fitting of UV, optical and mid-IR bands \citep{2007ApJS..173..267S,2018ApJ...859...11S}. With this approach, the dynamic range of sSFRs for central spiral galaxies becomes narrower, especially for galaxies with lower sSFR, as discussed in detail in \citet{2019ApJ...884L..52Z} and \citet{2024ApJ...973L..23D}; however, the general trends remain consistent.

The host dark matter halo mass ($M_{\rm halo}$) of each central galaxy is taken from \citet{2025ApJ...979...42Z}, who developed new machine learning (ML) models using 25 observable galaxy or group properties. These include the stellar mass of the central galaxy, the total stellar mass of the galaxies more massive than the mass completeness threshold in the same group, the total color matrix, group richness, and stellar age, SFR, color matrix of the central galaxy. Additionally, group halo masses are estimated separately for blue and red groups, as quenching processes can decouple the growth of the halos from that of the galaxies, complicating the mapping from group properties to halo mass. This innovative approach led to a notable improvement in the accuracy of halo mass measurements. By applying these models to observation data in the SDSS DR7 group catalog of \citet{Yang:2007}, they obtained accurate measurements of the $M_{\rm halo}$ for SDSS groups, down to $10^{11.5} \Msol$ or even lower. The derived stellar-to-halo mass relations (SHMR) for both blue and red central galaxies are in excellent agreement with those measured from weak lensing \citep[e.g.,][]{2006MNRAS.368..715M,2016MNRAS.457.4360Z,2018ApJ...862....4L,2021A&A...653A..82B}. However, the SHMRs for red and blue centrals, when derived using the widely-adopted halo masses estimated through the abundance matching (AM) technique, do not align with the weak lensing measurements. As demonstrated in \citet{2025ApJ...979...42Z}, the halo masses derived from the new ML approach show significant differences compared to those obtained using the AM technique, particularly around the Schechter stellar mass, $M_* \sim 10^{10.5} \Msol$. This corresponds to the ``golden halo mass" of approximately $10^{12} \Msol$ \citep{2019arXiv190408431D}, where the discrepancy exceeds 0.3 dex. Furthermore, using the improved ML-derived halo masses, the resulting halo mass function shows excellent agreement with theoretical predictions.

Since the star formation in the local universe primarily occurs in galactic disks, the size of the
disk component (R$_{\rm disk}$) — closely related to the properties of the dark matter (DM) halo, particularly its mass, spin, and concentration — serves as a key indicator of gas density. The $r$-band R$_{\rm disk}$ is taken from \citet{Simard2011}, who perform two-dimensional, point-spread-function-convolved, bulge + disk decompositions in the $g$ and $r$ bandpasses on a sample of over 1 million galaxies from the Legacy area of the SDSS DR7, utilizing a fitting model using a pure exponential disk and a de Vaucouleurs bulge.

\section{Results} \label{sec:results}

\subsection{Interplay between sSFR, SFE and $M_{\rm H_2}/M_{\rm HI}$}

It is well-known that in regular undisturbed spiral galaxies, the atomic gas extends well beyond the stellar and molecular gas disks, often reaching 2-4 times their diameter. This extensive distribution of atomic gas plays a critical role in the dynamics of star formation across different scales within galaxies \citep[e.g.,][]{2002ApJ...569..157W,2008AJ....136.2846B,2014ApJ...790...27L,2017A&A...605A..18C}. Although there are disparities in the spatial scales of \hi\ gas, \htwo\ gas and star formation, the star formation efficiency of \hi\ gas (SFE$_{\rm HI}$ = SFR/$M_{\rm HI}$) quantifies the galaxy's capability to transform available cold \hi\ gas into stars. The inverse of SFE$_{\rm HI}$, known as the \hi\ gas depletion timescale ($\tau_{\rm HI}$), describes the expected time to deplete the current \hi\ gas reservoir at ongoing star formation rates. It is beneficial to express SFE$_{\rm HI}$ as the product of the star formation efficiency of \htwo\ (SFE$_{\rm H_2}$, defined as SFR/$M_{\rm H_2}$), and the molecular-to-atomic gas mass ratio, $M_{\rm H_2}/M_{\rm HI}$. This relationship can be represented as SFE$_{\rm HI}$ = SFE$_{\rm H_2} \times M_{\rm H_2}/M_{\rm HI}$, highlighting how the efficiency of star formation from atomic gas depends on both the efficiency of molecular gas and the relative abundance of molecular to atomic gas.

\begin{figure}[htbp]
    \begin{center}
       \includegraphics[width=0.5\textwidth]{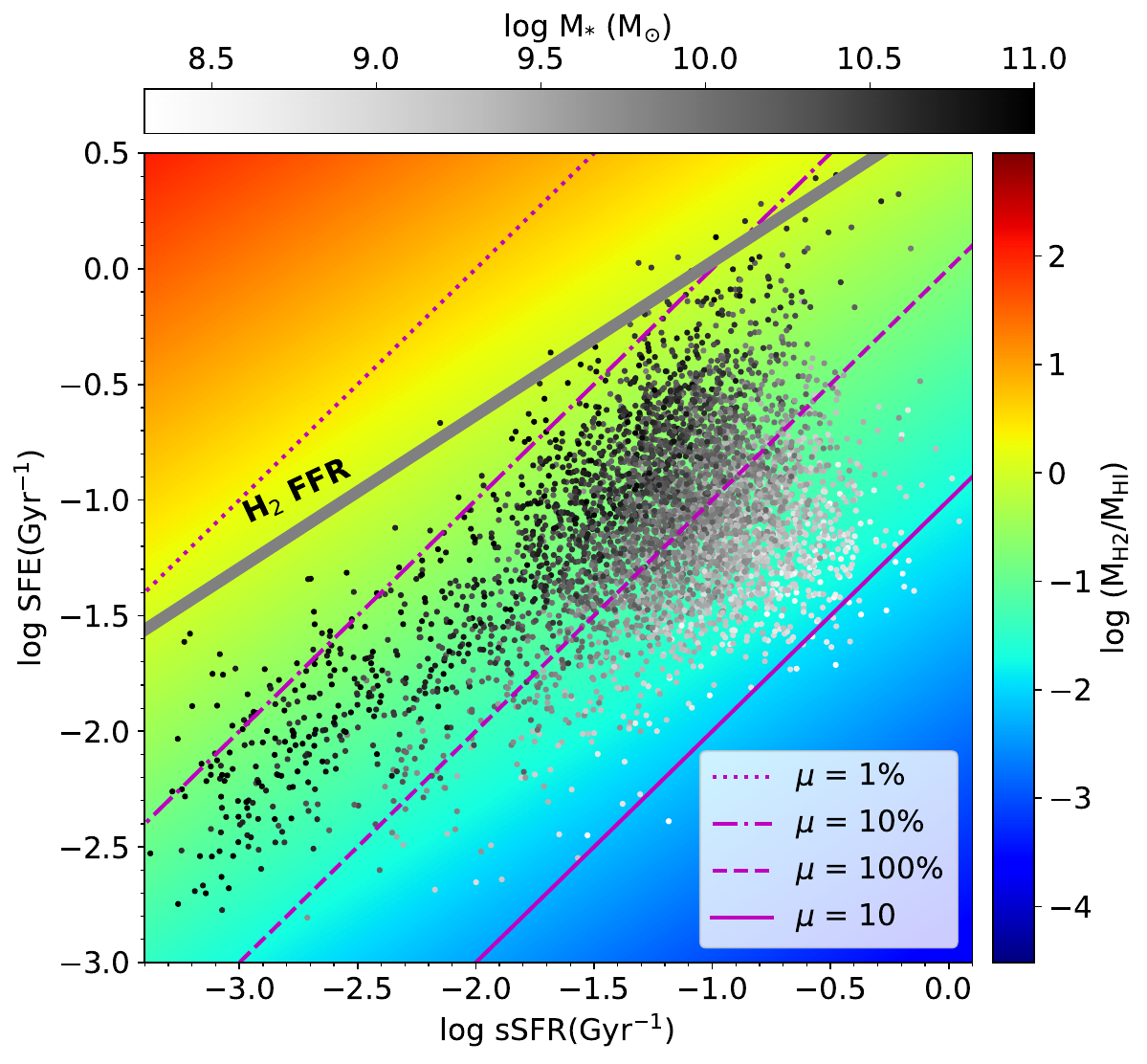}
    \end{center}
\caption{Distribution of galaxies on the SFE-sSFR plane. The dots show the location of central spiral galaxies in the ALFALFA-SDSS matched sample on the SFE$_{\rm HI}$-sSFR plane, with the grey scale representing the $M_*$. The purple diagonal lines indicate four different values of constant $\mu_{\rm HI}$ of 1\%, 10\%, 100\% and 10. The grey thick line indicates the best-fitted SFE$_{\rm H_2}$-sSFR relation found in \citet{2021ApJ...907..114D}, which is called the fundamental formation relation (FFR) of molecular gas. The background color represents the value of log (M$_{\rm H_2}$/M$_{\rm HI}$) estimated from the FFR. }
 \label{sfe-ssfr}
\end{figure}

Although SFE$_{\rm HI}$ and SFE$_{\rm H_2}$ have very different physical meanings, they share the same units (i.e., Gyr$^{-1}$). To better illustrate their distinct behaviors and provide a direct comparison, we place SFE$_{\rm HI}$ and SFE$_{\rm H_2}$ on the same plot. This approach effectively highlights the contrasting trends and dependencies of these two SFE measures, which are central to the analysis and interpretation of the results. The dots in Figure \ref{sfe-ssfr} show the distribution of central spiral galaxies from the ALFALFA-SDSS matched sample on the SFE$_{\rm HI}$-sSFR plane, with the grayscale of each dot indicating the galaxy's stellar mass ($M_*$). The purple diagonal lines denote four different constant values of the \hi\ to stellar mass ratio ($\mu_{\rm HI}$) of 1\%, 10\%, 100\% and 10. This figure shows a general positive correlation between SFE$_{\rm HI}$ and sSFR, though with significant scatter. Moreover, this relationship is systematically dependent on the $M_*$ for central spiral galaxies. At a given sSFR, more massive galaxies tend to exhibit larger SFE$_{\rm HI}$ and lower $\mu_{\rm HI}$ compared to low-mass systems, which is also shown in previous studies \citep[e.g.,][]{2012ApJ...756..113H,2017MNRAS.467.1083L,2020MNRAS.491.4843R,2024ApJ...973L..23D}. The $\mu_{\rm HI}$ of the sample spans a wide range, and low-mass dwarf galaxies potentially having an \hi\ gas fraction as high as approximately 90\% \citep[e.g.,][]{2018MNRAS.476..875C,2024ARA&A..62..113H}.

Additionally, the relationship between SFE$_{\rm H_2}$ and sSFR is delineated by the prominent thick grey line. As shown in \citet{2021ApJ...907..114D}, using data from the xCOLD GASS survey \citep{Saintonge:2017iz}, sSFR, SFE$_{\rm H_2}$, and molecular-to-stellar mass ratio $\mu_{\rm H_2}$ form a tight relationship. The scatter in this relation can be fully attributed to measurement errors in $\mu_{\rm H_2}$, SFE$_{\rm H_2}$ and sSFR, suggesting limited scope to further reduce the scatter by incorporating additional galaxy properties. Indeed, as further discussed and illustrated in \citep[e.g., Figure~2]{2021ApJ...915...94D}, this relation exhibits little to no dependence on other galaxy properties. 

Several well-known scaling relations, such as the integrated Kennicutt-Schmidt law, the star-forming main sequence (SF-MS), and the molecular gas main sequence, can all be derived from this relationship. Consequently, the SFE$_{\rm H_2}$ -$\mu_{\rm H_2}$ -sSFR relation is referred to as the Fundamental Formation Relation (FFR) of molecular gas. Galaxies across a range of stellar masses, sizes, structures, metallicities, and environments all evolve along this fundamental relation. This reflects that the star formation level in galaxies is primarily determined by the combined effects of galactic dynamical timescales (related to the gas depletion timescale, 1/SFE) and gas instability (associated with $\mu_{\rm H_2}$). These unique features make the molecular FFR an ideal framework to study galaxy formation and evolution. As shown in Figure \ref{sfe-ssfr}, the behavior of the \hi\ scaling relation contrast sharply with \htwo\ gas. The relation between SFE$_{\rm HI}$ and sSFR for central spiral galaxies exhibits significant scatter and is strongly dependent on the galaxy stellar mass, suggesting that \hi\ gas does not follow a similar FFR framework as \htwo\ gas.

The background color in Figure \ref{sfe-ssfr} represents the values of log (M$_{\rm H_2}$/M$_{\rm HI}$), as estimated from the ALFALFA survey and the \htwo\ FFR. Specifically, M$_{\rm H_2}$/M$_{\rm HI}$ is calculated as SFE$_{\rm HI}$/SFE$_{\rm H_2}$, where SFE$_{\rm H_2}$ is estimated from sSFR using the best orthogonal distance regression (ODR) fitted relation between SFE$_{\rm H_2}$ and sSFR: log SFE$_{\rm H_2}$ = 0.66 log sSFR + 0.69. This analysis follows the same ODR fitting method as in the upper right panel of Figure 6 in \citet{2021ApJ...907..114D}. It should be noted that we adopt the SFR from the value-added MPA-JHU catalog instead of the SFR estimates derived from the combination of mid-IR and UV data used in the previous study. As a result, the scaling relation obtained here is slightly different. SFE$_{\rm HI}$ is directed calculated from ALFALFA survey as SFE$_{\rm HI}$ = SFR/M$_{\rm HI}$. The typical uncertainty for SFR (in star-forming galaxies) and stellar mass is 0.2 dex and 0.1 dex, respectively, leading to an uncertainty in sSFR of approximately 0.22 dex. Based on the fitted relation, the uncertainty of SFE$_{\rm H_2}$ is about 0.15 dex. The typical uncertainty in log M$_{\rm HI}$ for the \hi\ detections in the ALFALFA-SDSS matched sample is approximately 0.06 dex, which combines uncertainties from both the \hi\ flux measurement and distance estimates \citep{2018ApJ...861...49H}. The resulting uncertainty of SFE$_{\rm HI}$ is 0.21 dex. Propagating these errors yields a total uncertainty in log (M$_{\rm H_2}$/M$_{\rm HI}$) of 0.26 dex.

The visualization in Figure \ref{sfe-ssfr} highlights the strong dependence of the M$_{\rm H_2}$/M$_{\rm HI}$ ratio on the $M_*$ and sSFR in central spirals, illustrating the dynamic interplay between gas composition and star formation activity. At a given sSFR, the $M_{\rm H_2}/M_{\rm HI}$ decreases with decreasing $M_*$, indicating a lower efficiency in converting \hi\ to \htwo\ in galaxies with lower masses. This reduced efficiency could be attributed to several factors, including lower metallicity and lesser gravitational binding energy in smaller galaxies, which are less conducive to the processes necessary for \htwo\ formation. 

Furthermore, as shown in Figure \ref{sfe-ssfr}, the slope of the molecular FFR (represented by the thick grey line, approximately 0.5 as indicated in \citeauthor{2021ApJ...907..114D} \citeyear{2021ApJ...907..114D}) differs from the constant $\mu_{\rm HI}$ lines (represented by the purple diagonal lines, with a slope of 1). This deviation indicates that a decrease in sSFR is correlated with a reduction in the $M_{\rm H_2}/M_{\rm HI}$ ratio, suggesting that galaxies with lower SFRs are also less efficient at converting \hi\ to \htwo. This trend could be attributed to quenching mechanisms, such as feedback processes or halo-related effects, that inhibit star formation.

It is interesting to note that the majority of \hi\ detections in the ALFALFA sample fall below the FFR line, with only a handful exceeding it. This suggests that SFE$_{\rm HI}$ is typically lower than SFE$_{\rm H_2}$, or equivalently, $\tau_{\rm HI}$ is greater than $\tau_{\rm H_2}$. For galaxies above the FFR, $\tau_{\rm HI}$ is less than $\tau_{\rm H_2}$, indicating that $M_{\rm HI}$ is less than $M_{\rm H_2}$ at a given sSFR. Since all the galaxies in our sample are central spiral galaxies that are expected to be less affected by environmental effect (e.g., \hi\ stripping), the relative lower $M_{\rm HI}$ compared to $M_{\rm H_2}$ suggests an enhanced \hi\ to \htwo\ conversion rate that is faster than the \htwo\ consumption rate, possibly caused by the compaction process triggered by mergers or inflow for instance. As \hi\ gas, the raw material for forming \htwo, is depleted faster than it can be replenished — also indicated by the low gas-to-mass ratio ($\mu \sim$ 10\%, as shown by the purple dot-dashed line in Figure \ref{sfe-ssfr}) — the \htwo\ gas will also deplete quickly. This leads to an unsustainable state, which explains the scarcity of galaxies in this region above the FFR. While galaxies may temporarily occupy this region, they cannot sustain it and quickly exhaust their gas, eventually falling below the FFR line.

\subsection{Influence of halo mass on the SFE and M$_{\rm H_2}$/M$_{\rm HI}$ ratio}

\begin{figure*}[htbp]
    \begin{center}
       \includegraphics[width=180mm]{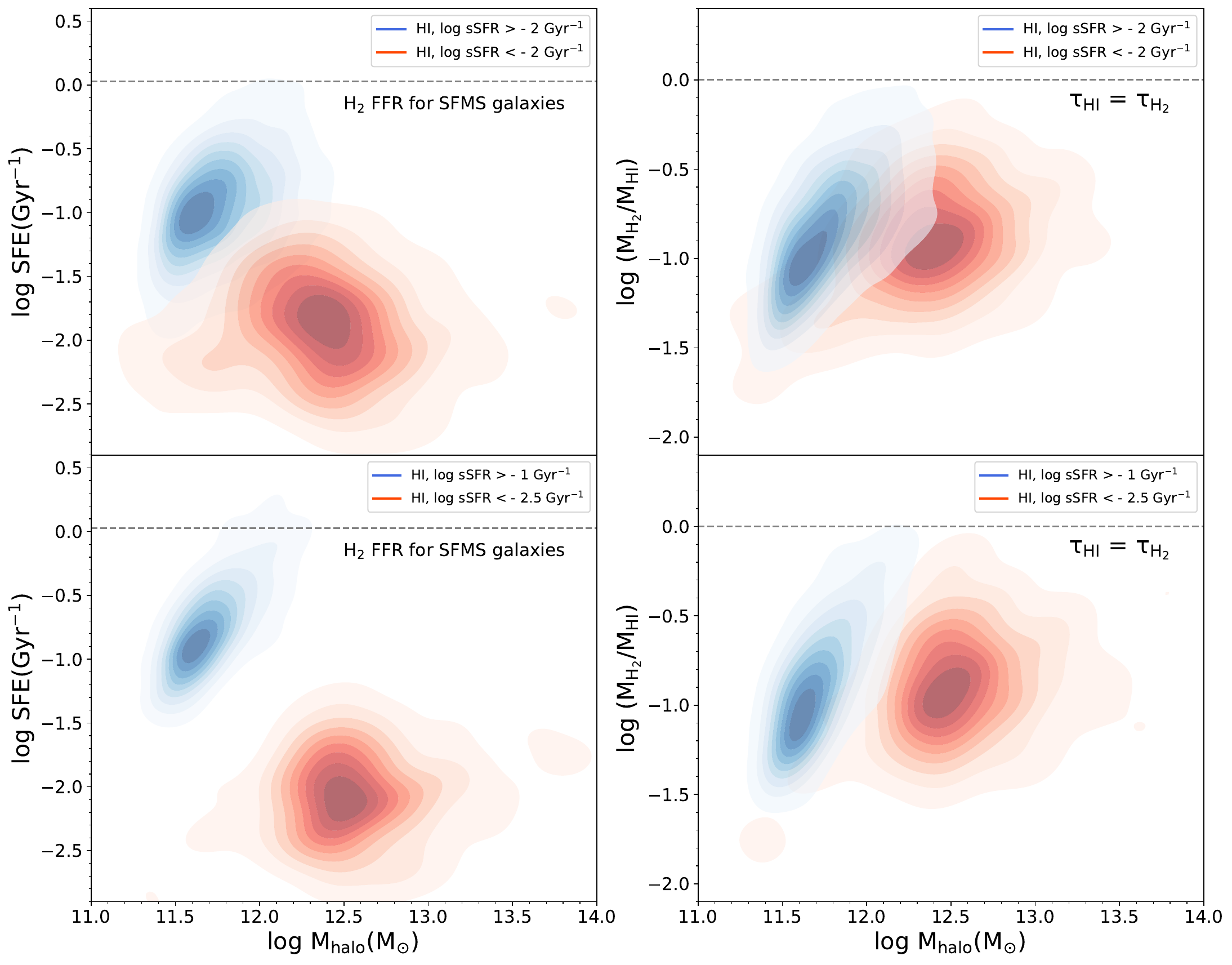}
    \end{center}
\caption{The SFE-M$_{\rm halo}$ relation (left panels) and (M$_{\rm H_2}$/M$_{\rm HI}$)-M$_{\rm halo}$ relation (right panels) for central spirals in the ALFALFA-SDSS matched sample. Level contours represent the number density of galaxies, with values normalized. Left: Blue contours show the SFE$_{\rm HI}$-M$_{\rm halo}$ relation for star-forming galaxies, while red contours represent passive galaxies. In the upper left panel, star-forming galaxies are defined by log sSFR $>$ -2 Gyr$^{-1}$, and red ones are those with log sSFR $<$ -2 Gyr$^{-1}$. In the lower left panel, a stricter criterion is applied: star-forming galaxies are defined by log sSFR $>$ -1 Gyr$^{-1}$ and passive galaxies by log sSFR $<$ -2.5 Gyr$^{-1}$. The horizontal dashed line indicates the molecular gas FFR for the star-forming main sequence galaxies. Right: The (M$_{\rm H_2}$/M$_{\rm HI}$)-M$_{\rm halo}$ relation is calculated using the molecular gas FFR. The same definitions for star-forming and passive galaxies are applied in the right panels as in the corresponding left panels. The horizontal dashed line indicates the position where the \hi\ depletion timescale is equal to that of \htwo. }
 \label{sfe-mhalo}
\end{figure*}

Figure \ref{sfe-mhalo} illustrates the SFE-M$_{\rm halo}$ relation (left panels) and (M$_{\rm H_2}$/M$_{\rm HI}$)-M$_{\rm halo}$ relation (right panels) for central spirals in the ALFALFA-SDSS matched sample. In the upper two panels, log sSFR $=$ -2 Gyr$^{-1}$ is used to distinguish star-forming and passive galaxies, while in the lower two panels, a more stringent definition is applied, with star-forming galaxies defined as log sSFR $>$ -1 Gyr$^{-1}$ and passive galaxies as log sSFR $<$ -2.5 Gyr$^{-1}$. Notably, the behaviors of star-forming and passive galaxies exhibit significant differences under both definitions.  

In the left panels, star-forming central spirals, represented by blue contours, show an upward trend in SFE$_{\rm HI}$ with increasing halo mass. This trend is especially pronounced in the lower panels, where the stricter definition of star-forming galaxies results in reduced scatter. This suggests that the halo mass plays a crucial role in regulating the conversion efficiency from \hi\ gas to stars within these galaxies and the conversion of \hi\ gas to stars is more efficient in more massive halos. This trend contrasts with the behavior of molecular gas. \citet{2021ApJ...907..114D} highlights that while SFE$_{\rm H_2}$ is closely linked with sSFR, this relation shows no dependence on other galaxy properties, including halo mass. This independence is depicted by a horizontal dashed line in the left panel, indicating that SFE$_{\rm H_2}$ remains consistent across different halo masses for a given sSFR. Also, the blue contours are on average lower than the dashed line, indicating that the SFE$_{\rm HI}$ is typically lower than SFE$_{\rm H_2}$, as also shown in Figure \ref{sfe-ssfr}.

The right panels illustrate the relationship between halo mass and the efficiency of converting \hi\ to \htwo\ (M$_{\rm H_2}$/M$_{\rm HI}$). For star-forming galaxies, the conversion process from \hi\ to \htwo\ also becomes significantly more efficient in more massive halos, ultimately achieving the equilibrium where the depletion timescales for \hi\ and \htwo\ are equal, illustrated by the gray dashed line. It also shows that most galaxies have M$_{\rm H_2} <$ M$_{\rm HI}$, which is consistent with SFE$_{\rm HI} <$ SFE$_{\rm H_2}$, or equivalently, $\tau_{\rm HI} >$ $\tau_{\rm H_2}$, as illustrated in Figure \ref{sfe-ssfr}.

Since dark matter (DM) does not directly contribute significantly to the mid-plane pressure, its contribution is often neglected in the vast majority of previous studies, where baryonic components are considered the dominant contributors. However, DM can indirectly influences mid-plane pressure by altering the vertical distribution of stars and gas. Indeed, several theoretical studies \citep{2016MNRAS.462.3053B,2020A&A...638A..66P,2020MNRAS.498.3664G} have highlighted the significant impact of DM halos on mid-plane pressure in spiral galaxies. According to these studies, the gravitational potential of the DM halo contributes to the overall vertical stability and structure of the galactic disk. Ignoring the gravitational potential of the DM halo can lead to a significant underestimation of the vertical baryon density near the disk mid-plane. This underestimation, in turn, results in lower mid-plane pressures, which are crucial for efficiently converting \hi\ into \htwo\ and promoting star formation.

Theoretically, the stronger gravitational potential provided by DM causes gas to settle more tightly within the disk, increasing its surface density \citep[e.g.,][]{2012ApJ...748..101S}. This increased gas surface density leads to a rise in local mid-plane pressure. By enhancing the vertical compression of gas, DM indirectly improves the efficiency of \hi-to-\htwo\ conversion \citep[e.g.,][]{2010MNRAS.409..515F}. The higher mid-plane pressure supports the formation of molecular clouds, which are essential for star formation. In this way, DM indirectly promotes the conditions necessary for higher \htwohi, thereby boosting star formation efficiency.

Higher mid-plane pressure, often associated with more massive DM halos, enhances the conditions required for the formation of molecular clouds. As a result, galaxies embedded in more massive DM halos should have a higher \htwohi\ ratio, reflecting a more efficient conversion of \hi\ to \htwo\ and leading to an increased SFR. This indirect impact of the DM halo is also expected to be particularly crucial in low-mass galaxies, where the baryonic mass (gas and stars) alone may not provide sufficient gravitational potential to stabilize the vertical structure. In these low-mass galaxies, the gravitational influence of the DM halo becomes critical for maintaining the disk's vertical equilibrium, supporting higher mid-plane pressure, and promoting more efficient \hi-to-\htwo\ conversion and star formation.

These theoretical perspectives all align well with our observational results, as shown in Figure \ref{sfe-mhalo} and \ref{sfe-ssfr_2}, which underscore the critical role of DM halos as global regulators of galaxy-wide star formation, a factor that may have been largely underappreciated in previous studies.

As shown in both panels in Figure \ref{sfe-mhalo}, when the halo mass exceeds $10^{12} \Msol$, galaxies are predominantly quenched, as indicated by the red contours. At a given halo mass, this quenching is accompanied by lower SFE$_{\rm HI}$ and a reduced ratio of M$_{\rm H_2}$/M$_{\rm HI}$. This is consistent with previous results from \citet{2019ApJ...884L..52Z} and \citet{2024ApJ...973L..23D}, which show that during the quenching process, the amount of \hi\ gas in central spiral galaxies remains largely unchanged, whereas the amount of \htwo\ gas strongly decreases, leading to a lower M$_{\rm H_2}$/M$_{\rm HI}$ ratio and lower SFE$_{\rm HI}$. It is interesting to note that, on average, the passive galaxy population exhibits higher M$_{\rm H_2}$/M$_{\rm HI}$ ratios compared to the star-forming galaxy population. However, the average halo mass (and also stellar mass) of passive galaxies is significantly higher than that of star-forming galaxies. These represent very different populations: for example, local star-forming galaxies with halo masses around $\sim$ $10^{11.6} \Msol$ (the peak of the blue contours) are not progenitors of the passive galaxies with halo masses around $\sim 10^{12.5} \Msol$ (the peak of the red contours). Therefore, a more meaningful comparison of the M$_{\rm H_2}$/M$_{\rm HI}$ ratios should be made at similar halo or stellar masses. We will discuss in detail in Figure \ref{model} later that evolutionary paths \ding{172} and \ding{173} are more physically plausible than path \ding{174}.

These findings suggest that more massive halos, specifically those with masses greater than $10^{12} \Msol$, significantly influence the quenching of galaxies by inhibiting the conversion of gas into stars and suppressing star formation activities. This is consistent with the halo quenching theory proposed by \citet{2006MNRAS.368....2D}. According to this theory, ones a halo reaches a critical mass, it transitions from cold accretion to hot accretion and it can retain hot gas in a virialized state. This phenomenon and additional heating source such as AGN feedback effectively halts new star formation in the galaxy, transitioning it into a quenched state. 

This phenomenon is crucial in understanding the evolution of galaxies, as it explains why some galaxies cease forming new stars and transition to a passive state. Additionally, this process highlights the importance of halo mass in regulating star formation and the overall growth of galaxies.

\subsection{Impact of disk size on the SFE$_{\rm HI}$ and M$_{\rm H_2}$/M$_{\rm HI}$ ratio}

\begin{figure*}[htbp]
    \begin{center}
       \includegraphics[width=180mm]{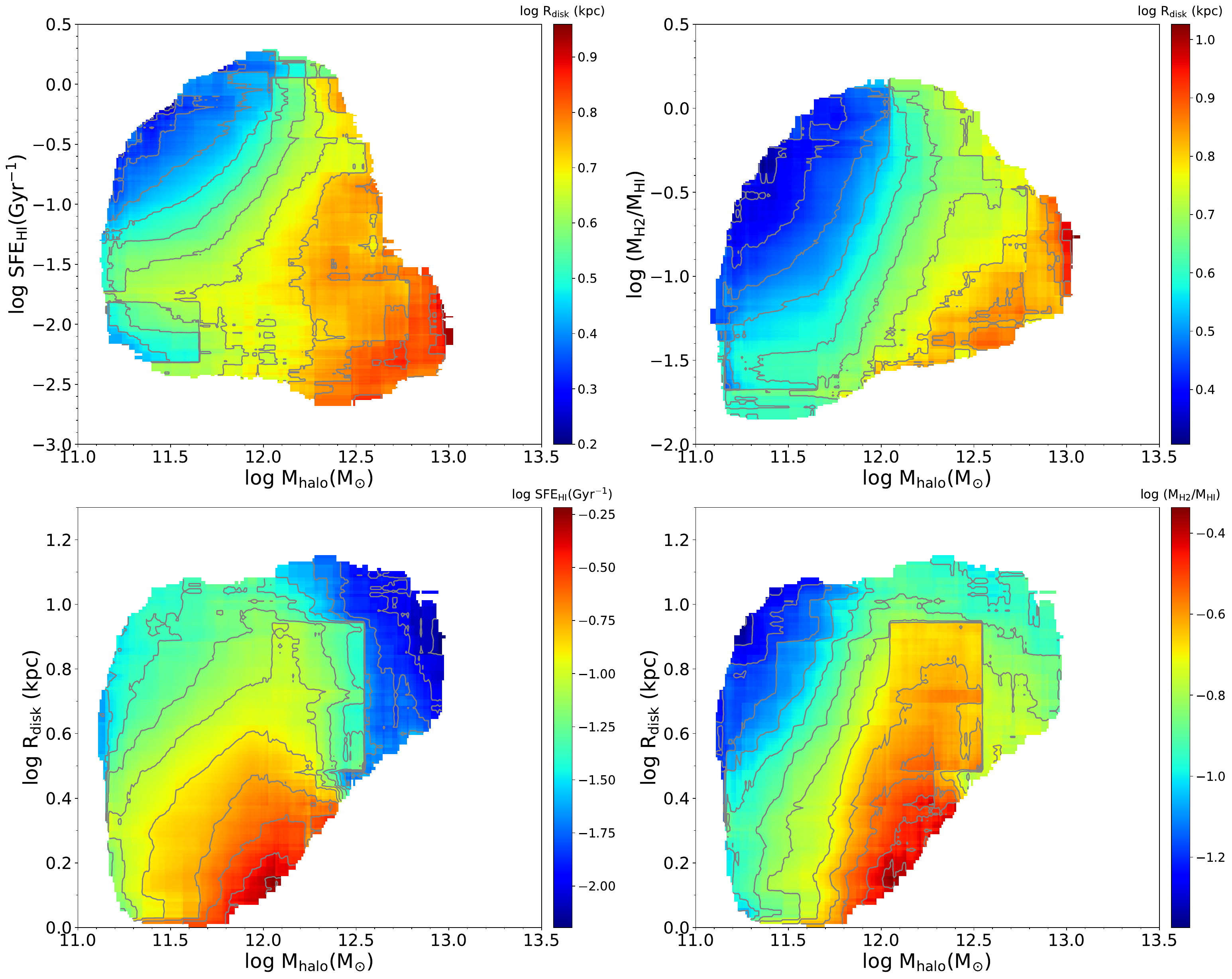}
    \end{center}
\caption{Upper: the average disk radius (log R$_{\rm disk}$) on the SFE$_{\rm HI}$-M$_{\rm halo}$ plane (upper left panel) and (M$_{\rm H_2}$/M$_{\rm HI}$)-M$_{\rm halo}$ plane (upper right panel) for central spirals in the ALFALFA-SDSS matched sample. Lower: the average log SFE$_{\rm HI}$ (lower left panel) and log (M$_{\rm H_2}$/M$_{\rm HI}$) (lower right panel) on the R$_{\rm disk}$-M$_{\rm halo}$ plane. All panels are obtained by using a moving box of size 0.5 dex in both x-axis and y-axis.}
 \label{figure3}
\end{figure*}

In the local universe, star formation predominantly occurs in galactic disks. In the theoretical framework, the size of these disks (R$_{\rm disk}$), which we use as a proxy for gas density, is primarily determined by the properties of the dark matter (DM) halo, especially its mass, spin and concentration. In general, a larger halo mass, higher spin and lower concentration lead to larger disk sizes \citep[e.g.,][]{1998MNRAS.295..319M,2001MNRAS.321..559B,2013ApJ...764L..31K,2019MNRAS.488.4801J,2022ApJ...937L..18D,2023MNRAS.518.1002R,2024arXiv240314749L}. In addition to halo properties, various baryonic processes, such as angular momentum transfer \citep[e.g.,][]{1980MNRAS.193..189F,2015MNRAS.449.2087D}, gas accretion \citep[e.g.,][]{2006MNRAS.368....2D}, merger history \citep[e.g.,][]{2012MNRAS.423.1544S}, and feedback mechanisms \citep[e.g.,][]{2010Natur.463..203G}, also play essential roles in shaping disk size.

The upper panels in Figure \ref{figure3} illustate the average log R$_{\rm disk}$ on the SFE$_{\rm HI}$-M$_{\rm halo}$ plane (upper left panel) and (M$_{\rm H_2}$/M$_{\rm HI}$)-M$_{\rm halo}$ plane (upper right panel), calculated with a moving box of 0.5 dex in M$_{\rm halo}$, R$_{\rm disk}$ and M$_{\rm H_2}$/M$_{\rm HI}$. At a given DM halo mass, both SFE$_{\rm HI}$ and M$_{\rm H_2}$/M$_{\rm HI}$ show a strong correlation with the disk size. Specifically, at a given halo mass, larger disks tend to have lower SFE$_{\rm HI}$ and reduced M$_{\rm H_2}$/M$_{\rm HI}$. This trend is particularly evident for halos with masses below $10^{12} \Msol$, where cold accretion mode dominates the gas supply. In the lower panels, the log SFE$_{\rm HI}$ and log (M$_{\rm H_2}$/M$_{\rm HI}$) values are color-coded on the R$_{\rm disk}$-M$_{\rm halo}$ plane, revealing consistent results with those observed in the upper panels.

These findings suggest that disk size, primarily regulated by the spin and concentration of the DM halo, significantly influences gas availability and star formation within galaxies. Specifically, DM halos with higher angular momentum (i.e., larger spin parameters) and lower concentration values tend to host more extended disks. In such galaxies, the gas is likely distributed over a larger area, potentially leading to lower gas densities. This decreased density can reduce the molecular gas fraction, resulting in a lower M$_{\rm H_2}$/M$_{\rm HI}$ and SFE$_{\rm HI}$, as less dense gas is less likely to collapse and form stars. The relationship between halo properties and disk size thus implies that fundamental characteristics of the halo, like spin and concentration, indirectly control the conditions necessary for efficient star formation. Consequently, variations in halo spin and concentration could lead to significant diversity in star formation rates and gas compositions across galaxies.

Moreover, the mass of the DM halo contributes significantly to this relationship, as shown in Figure \ref{figure3}. The virial radius, which represents the boundary within which the halo is in approximate virial equilibrium, scales directly with halo mass and affects the overall gravitational potential. Larger DM halos, with greater virial radii, create deeper gravitational potential wells, which support more extended disks. This allows the gas within these halos to be distributed over a larger volume, often resulting in lower gas densities that can reduce the efficiency of star formation within the galaxy. 

Therefore, the combined effects of halo spin, concentration, and mass collectively determine disk size, which in turn influence the gas density, star formation efficiency, and gas composition. Understanding these dependencies is essential for comprehending galaxy formation and evolution, as it reveals how dark matter halo characteristics fundamentally contribute to the overall growth and activity of galaxies.

\subsection{Results at different stellar mass bins}

\begin{figure*}[htbp]
    \begin{center}
       \includegraphics[width=180mm]{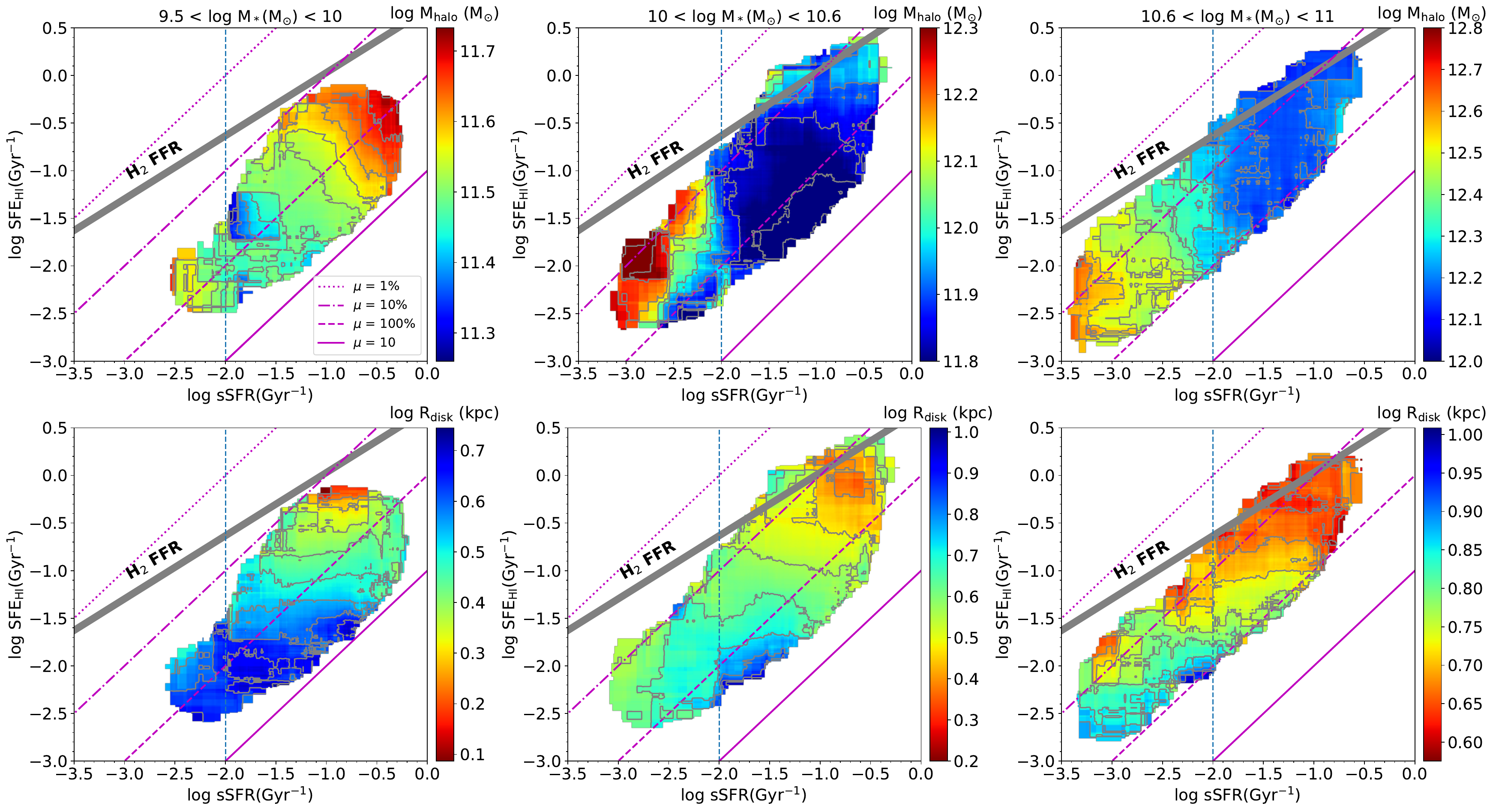}
    \end{center}
\caption{The average halo mass M$_{\rm halo}$ (upper panels) and disk radius R$_{\rm disk}$ (lower panels) on the SFE$_{\rm HI}$-sSFR plane for central spirals in the ALFALFA-SDSS matched sample, obtained by using a moving box of size 0.5 dex in sSFR and 0.5 dex in SFE$_{\rm HI}$. The galaxies are divided into low-mass bin 9.5 $<$ log M$_* (\Msol) <$ 10 (left panels), intermediate-mass bin 10 $<$ log M$_* (\Msol) <$ 10.6 (middle panels) and high-mass bin 10.6 $<$ log M$_* (\Msol) <$ 11 (right panels). In all panels, the blue vertical line indicates the approximate position between star-forming galaxies and passive galaxies. The purple diagonal lines indicate four different values of constant $\mu_{\rm HI}$ of 1\%, 10\%, 100\% and 10. The grey thick line indicates the molecular gas fundamental formation relation (FFR) found in \citet{2021ApJ...907..114D}.}
 \label{sfe-ssfr_2}
\end{figure*}

Stellar mass is a fundamental property that influences local gravity and correlates with nearly all galaxy properties. Given the strong correlation between halo mass and stellar mass, particularly for star-forming galaxies, we analyze the ALFALFA-SDSS sample by dividing it into three fixed stellar mass bins to help disentangle the influences of M$_{\rm halo}$ and M$_*$, as shown in Figure \ref{sfe-ssfr_2}. The blue vertical line in each panel indicates the approximate boundary (log sSFR = -2 Gyr$^{-1}$) that separates star-forming galaxies from passive ones. The purple diagonal lines indicate four different values of constant $\mu_{\rm HI}$ of 1\%, 10\%, 100\% and 10. The grey thick line indicates the molecular gas fundamental formation relation (FFR) proposed in \citet{2021ApJ...907..114D}. This figure illustrates the relationships among \hi\ gas, \htwo\ gas and star formation activity.

The upper panels of Figure \ref{sfe-ssfr_2} show the average DM halo mass (M$_{\rm halo}$), while the lower panels show the average disk radius (R$_{\rm disk}$) on the SFE$_{\rm HI}$-sSFR plane, calculated using a moving box of 0.5 dex in sSFR and SFE$_{\rm HI}$. For lower mass galaxies (9.5 $<$ log $M_* (\Msol) <$ 10), as shown in the left panels, the SFE$_{\rm HI}$ and the conversion from \hi\ to \htwo\ increase with higher halo mass and smaller disk radius, suggesting that DM halo mass, halo spin and concentration play a significant role in regulating the available gas for star formation in low-mass systems.

The halo masses of the most massive galaxies (10.6 $<$ log $M_* (\Msol) <$ 11) are mostly exceeds $10^{12} \Msol$. As shown in the upper right panel, when the halo mass of the galaxy exceeds $10^{12} \Msol$, the average sSFR falls below 0.01 Gyr$^{-1}$, indicating that these galaxies are quenched. In these massive galaxies, DM halo mass plays a crucial role in governing the onset of quenching, consistent with the halo quenching scenario. The lower right panel shows that smaller disk size elevates the SFE$_{\rm HI}$ and the conversion from \hi\ to \htwo, underscoring the impact of halo spin and concentration on gas dynamics and subsequent star formation processes.

Galaxies of intermediate stellar masses (10 $<$ log $M_* (\Msol) <$ 10.6) display the combined regulatory influences of both halo mass and disk size. These galaxies show that the halo begins to exert a quenching effect once the mass reaches a critical threshold, typically around $10^{12} \Msol$, marking a transition in the galaxy's evolution characterized by reduced star formation activity.

Overall, Figure \ref{sfe-ssfr_2} provides a comprehensive view of how various properties of dark matter halos, including spin, concentration, and mass, collectively influence galaxy evolution. Halo spin and concentration play a crucial role in determining internal galactic structure, such as disk size, which regulates the conversion of \hi\ to \htwo\ and affects star formation rates across different stellar masses. Meanwhile, halo mass perform differently across different stellar masses. In low-mass galaxies (where most halos have masses less than $10^{12} \Msol$), the dark matter halo promotes the conversion of \hi\ to \htwo\ gas primarily through cold accretion mode, in which gas flows smoothly into the galaxy and sustains star formation. In this regime, halo-driven quenching has not yet begun to operate. In massive galaxies (log M$_* (\Msol) >$ 10.6), the halo transitions to a hot accretion mode, where the gas is shock-heated upon entry into the halo. Here, the halo effectively acts as a quenching switch, inhibiting further star formation.

\section{A Plausible Halo-Regulated Evolutionary Scenario}

\begin{figure*}[htbp]
    \begin{center}
       \includegraphics[width=150mm]{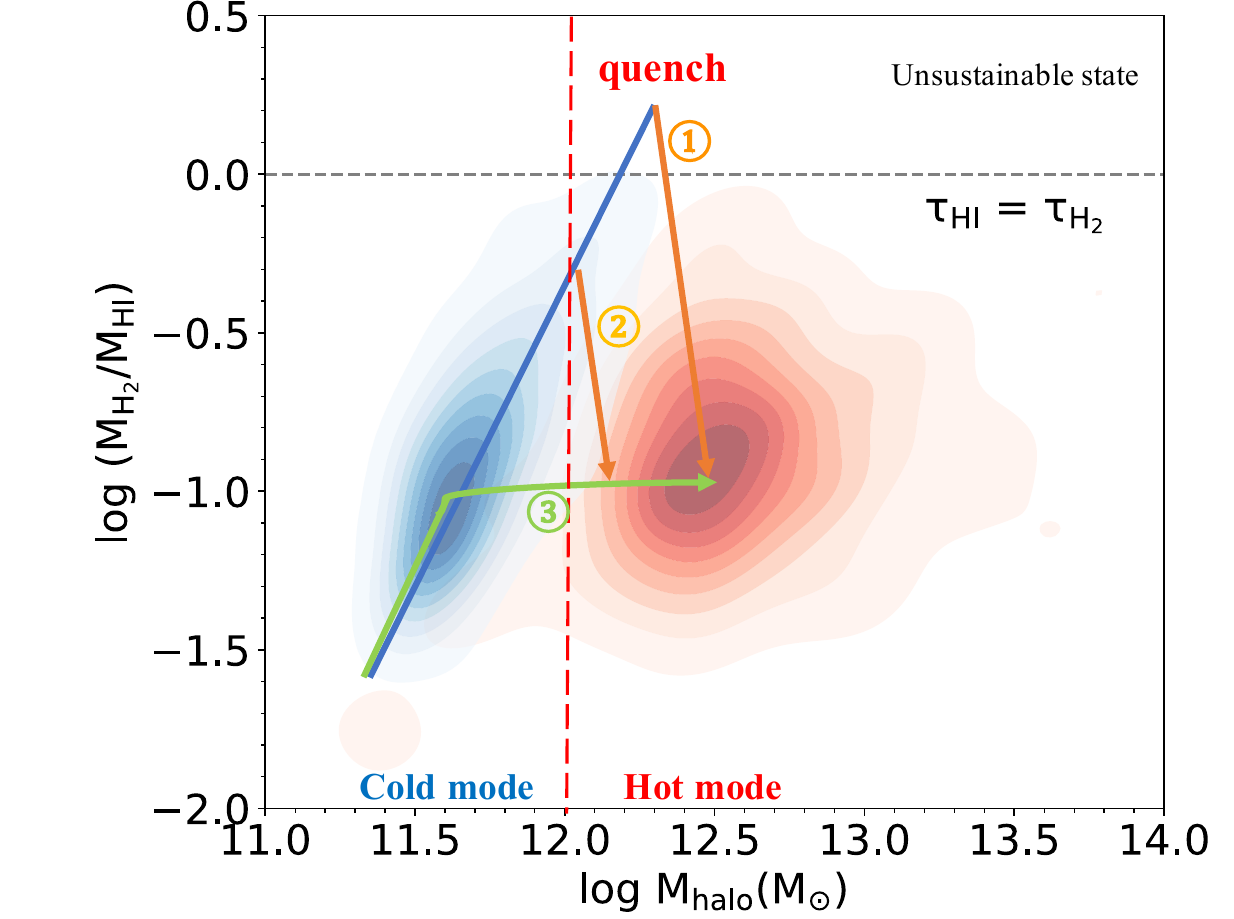}
    \end{center}
\caption{Proposed evolutionary scenario illustrating how halo mass growth regulates the conversion of \hi\ to \htwo, star formation, and the eventual quenching of galaxies. Based on the lower right panel of Figure \ref{sfe-ssfr}, this diagram includes possible evolutionary paths (labeled as \ding{172}\ding{173} and \ding{174}) that galaxies may follow as they transition through different states. In lower-mass halos, galaxies are supported primarily in a cold accretion mode, which promotes efficient star formation. As halo mass increases, galaxies progress along paths \ding{172} (quenching at higher redshift) and \ding{173} (quenching at low redshift), with gas transitioning to a hot accretion mode that inhibits the conversion of \hi\ to \htwo, suppressing star formation and eventually leading to a quenched state.}
 \label{model}
\end{figure*}

To better understand the role of the dark matter halo in driving star formation, we illustrate the plausible evolutionary paths of galaxies on the M$_{\rm H_2}$/M$_{\rm HI}$–M$_{\rm halo}$ plane. As shown in Figure \ref{model}, in principle, there are two distinct paths (labeled as \ding{172}\ding{173} and \ding{174}) by which galaxies can progress to join the cloud of quenched galaxies in massive halos. Due to the rapid decline in specific halo accretion rates over cosmic time, halos can only significantly increase their mass at high redshifts, with limited mass growth at z $<$ 1 \citep[e.g.,][]{2014MNRAS.443.3643P}. Star-forming galaxies form a relatively tight, single sequence on this plane, without any significant bending at higher halo masses. This suggests that the more likely paths are routes \ding{172} and \ding{173}, where star-forming galaxies evolve along the relatively tight M$_{\rm H_2}$/M$_{\rm HI}$–M$_{\rm halo}$ sequence specific to star-forming galaxies.

The evolutionary path \ding{172} leads galaxies to cross above the horizontal dashed line, entering an unsustainable state where \htwo\ gas becomes more dominant than \hi\ gas. In this state, galaxies have an elevated \htwo\ conversion efficiency within massive halos, resulting in high star formation rates and short gas depletion timescales. This is also the phase where galaxies are most susceptible to quenching. Once any mechanism disrupts the supply of \hi\ gas, the galaxy rapidly becomes quenched as gas is depleted by star formation (note that in this state, the galaxy has a shorter \hi\ depletion timescale compared to \htwo, often less than $\sim$ 1 Gyr, as evidenced in the left panels of Figure \ref{sfe-mhalo}, the horizontal dashed line).

Conversely, star-forming galaxies with lower M$_{\rm H_2}$/M$_{\rm HI}$ and SFE$_{\rm HI}$ values, such as those at the peak of the blue sequence distribution indicated by contour lines in Figure \ref{sfe-mhalo}, exhibit M$_{\rm H_2}$/M$_{\rm HI}$ $\sim$ 0.1 and SFE$_{\rm HI}$ $\sim$ 0.1 Gyr$^{-1}$. These galaxies contain roughly ten times more \hi\ than \htwo\ gas and have long \hi\ depletion timescales of $\sim$ 10 Gyr, which is comparable to the Hubble timescale. As a result, these galaxies are less prone to quenching through gas depletion by star formation alone, implying that an additional mechanism is required to facilitate the quenching process.

As the halo mass continues to grow, it eventually surpasses halo mass threshold for shock heating and entering hot mode \citep[e.g.,][]{2003MNRAS.345..349B,2006MNRAS.368....2D,2005MNRAS.363....2K}. Quenching then likely occurs, might aided by AGN feedback \citep[e.g.,][]{2006MNRAS.365...11C,2012ARA&A..50..455F}. At this point, galaxies evolve almost vertically downward on the plot, with only modest increases in halo mass towards z $=$ 0, as illustrated by route \ding{172}. It should be noted that the threshold halo mass for transitioning from cold mode to hot mode accretion increases with redshift. This explains why quenched galaxies are concentrated in the high-mass region (M$_{\rm halo}$ $\sim 10^{12.5} \Msol$), which is significantly more massive than where present-day star-forming galaxies are typically found. This also explains why, in the local universe, few galaxies cross above the horizontal dashed line to reach the unsustainable \htwo-dominated state, as the shock heating halo mass threshold at low redshift is around $10^{12} \Msol$, leading galaxies to evolve along route \ding{173}.

Future SKA observations of \hi\ in high-redshift galaxies will provide a direct test of this evolutionary scenario.

\section{Summary} \label{sec:summary}

In this study, we explore the physical mechanisms that drive star formation, a crucial factor in advancing our understanding of galaxy evolution. Due to the sensitivity of neutral hydrogen (\hi) to external processes, we focus exclusively on central spiral galaxies. Using the ALFALFA-SDSS matched sample, with careful joint incompleteness corrections for both the ALFALFA and SDSS samples, and the Fundamental Formation Relation (FFR; \citealt{2021ApJ...907..114D,2021ApJ...915...94D}) derived from local \htwo\ observations, we investigate the interrelationships between key galaxy properties related to star formation, with particular emphasis on the impact of dark matter halos, using the improved halo masses derived by \citet{2025ApJ...979...42Z}. The main results are summarized as follows:

(1) M$_{\rm H_2}$/M$_{\rm HI}$ strongly depends on stellar mass and sSFR. A decrease in sSFR is correlated with a reduction on the $M_{\rm H_2}/M_{\rm HI}$ ratio, suggesting that galaxies with lower SFRs are less efficient at converting \hi\ to \htwo. In the SFE–sSFR plane, most galaxies fall below the \htwo\ FFR, with SFE$_{\rm HI}$ being consistently lower than SFE$_{\rm H_2}$ (equivalently, $\tau_{\rm HI}$ is larger than $\tau_{\rm H_2}$). Above the FFR, \htwo\ is depleted faster than it can be replenished, leading to an unsustainable star formation state, which explains the scarcity of galaxies in this region (Figure \ref{sfe-ssfr}).

(2) For star-forming galaxies, both SFE$_{\rm HI}$ and M$_{\rm H_2}$/M$_{\rm HI}$ increase rapidly and monotonically with halo mass (Figure \ref{sfe-mhalo}), indicating a higher efficiency in converting \hi\ to \htwo\ in more massive halos. This trend ultimately leads to an unsustainable state where SFE$_{\rm HI}$ exceeds SFE$_{\rm H_2}$. For halos with masses exceeding $10^{12} \Msol$, galaxies predominantly experience quenching. These trends also hold across different fixed stellar mass bins (Figure \ref{sfe-ssfr_2}).

The role of DM halos as global regulators of star formation has been largely neglected in many previous studies, for instance, their impact on mid-plane pressure. While DM does not directly contribute significantly to the mid-plane pressure, it indirectly influences the mid-plane pressure by altering the vertical distribution of stars and gas. Some theoretical calculations \citep[e.g.,][]{2020A&A...638A..66P} suggest that ignoring the DM halo potential can indeed significantly underestimate the vertical density close to the mid-plane, thereby considerably affecting the mid-plane pressure. This theoretical perspective aligns with our observational results.

(3) At a given halo mass (Figure \ref{figure3}) or a given stellar mass (Figure \ref{sfe-ssfr_2}), smaller disks tend to have higher SFE and higher M$_{\rm H_2}$/M$_{\rm HI}$. This suggests that halo mass, spin and concentration, which regulate the disk size, also significantly influences \hi\ to \htwo\ conversion and star formation within galaxies. 

(4) We proposed a plausible evolutionary scenario on the M$_{\rm H_2}$/M$_{\rm HI}$–M$_{\rm halo}$ plane (Figure \ref{model}) to illustrate how halo mass growth regulates the conversion of \hi\ to \htwo, star formation, and the eventual quenching of galaxies. At high redshifts, halos grow rapidly, driving galaxies along a tight sequence where the \hi\ to \htwo\ conversion increases in more massive halos. This evolution may push galaxies into an unsustainable state dominated by \htwo, with elevated star formation rates and rapid gas depletion, making them more susceptible to quenching once the \hi\ supply is disrupted. As halo mass surpasses a critical threshold for shock heating, hot mode accretion begins, suppressing the \hi\ supply. With the aid of AGN feedback, the galaxy will be quenched as gas is depleted by star formation. This threshold, which increases with redshift, explains the concentration of quenched galaxies in massive halos. Future high-redshift \hi\ observations with SKA will test this scenario.

These findings underscore the critical role of DM halos — specifically their mass, spin, and concentration — as global regulators of galaxy-wide star formation. This influence is exerted through various mechanisms, such as impacting the vertical baryon distribution which in turn affects the mid-plane pressure or through other halo-related processes. These factors may have been largely underappreciated in previous studies.


\section{Acknowledgments}

J.D. acknowledges the National Science Foundation of China (NSFC) Grant number 12303010. Y.P. acknowledges
the NSFC Grant numbers 12125301, 12192220 and 12192222, the science research grants from the China Manned Space Project with number CMS-CSST-2021-A07, and support from the New Cornerstone Science Foundation through the XPLORER PRIZE. Q.G. acknowledges the NSFC Grant numbers 12192222, 12192220 and 12121003. L.C.H. acknowledges the National Key R\&D Program of China (2022YFF0503401), the National Science Foundation of China (11991052, 12233001), and the China Manned Space Project (CMS-CSST-2021-A04, CMS-CSST-2021-A06). C.Z. was supported by the KASI-Yonsei Postdoctoral Fellowship and was supported by the Korea Astronomy and Space Science Institute under the R\&D program (Project No. 2025-1-831-02), supervised by the Korea AeroSpace Administration. D.L. acknowledges the NSFC Grant number 11988101. D.L. is a New Cornerstone investigator. F.Y. acknowledges the NSFC Grant number 12133008, 12192223, and 12361161601, and China Manned Space Project (CMS-CSST-2021-B02).


\appendix

\section{Mass - SFR and Mass - Size Relations for Central Spiral Galaxies}

\begin{figure*}[htbp]
     \begin{center}
       \includegraphics[width=180mm]{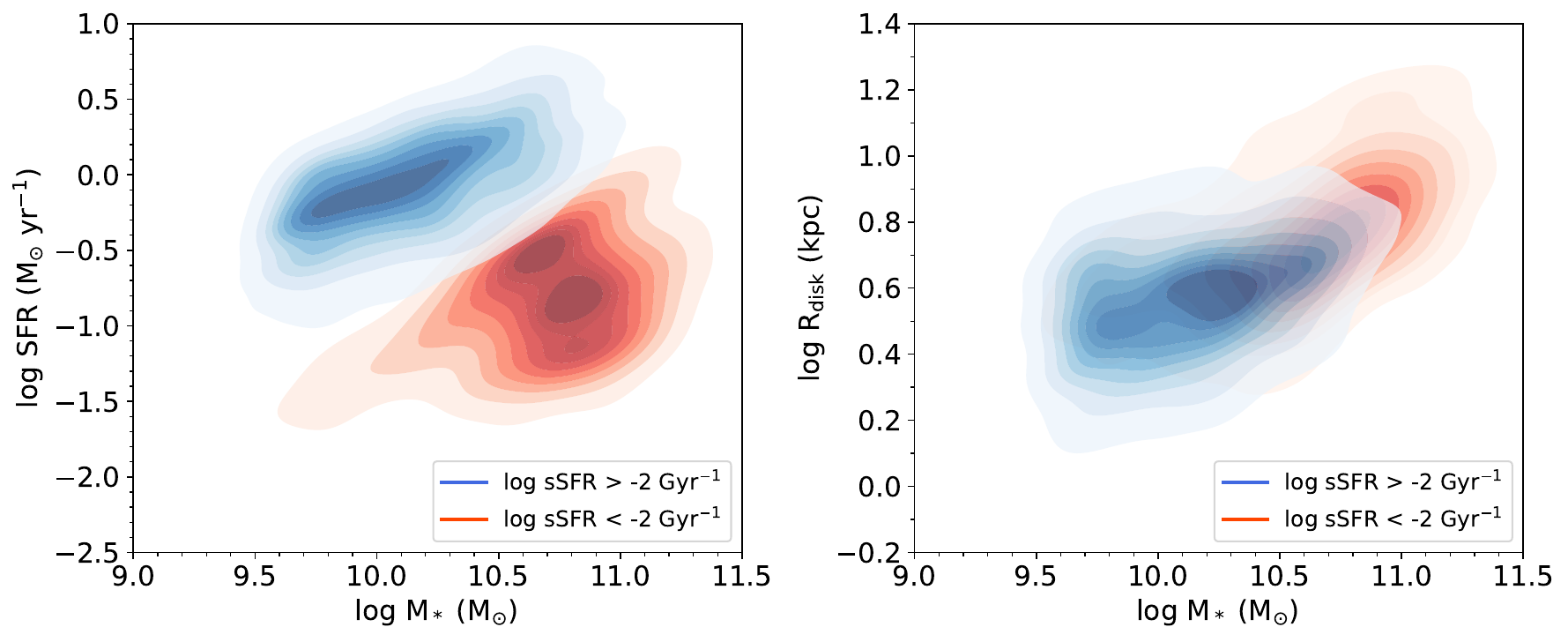}
    \end{center}
\noindent\textbf{Figure A1. }{Distribution of central spiral galaxies in the ALFALFA-SDSS matched sample on the star formation rate (SFR)–stellar mass (M$_*$) plane (left panel) and the disk size (R$_{\rm disk}$) –M$_*$ plane (right panel). Contours represent galaxy populations separated by sSFR: blue contours show galaxies with log sSFR $>$ -2 Gyr$^{-1}$, and red contours show galaxies with log sSFR $<$ -2 Gyr$^{-1}$. Both populations have been normalized.}
 \label{appendix}
\end{figure*}

Figure A1 shows the relationship between SFR and disk size with stellar mass for central spiral galaxies in the ALFALFA-SDSS matched sample, which are divided into two populations based on their sSFRs: galaxies with log sSFR $>$ -2 Gyr$^{-1}$ (blue contours) and those with log sSFR $<$ -2 Gyr$^{-1}$ (red contours). Both populations have been normalized.

The left panel of Figure A1 shows the distribution of central spiral galaxies on the SFR-M$_*$ plane. This distribution resembles the well-established structure observed in the general galaxy population, typically characterized by two main regions: the star-forming main sequence (SFMS) and the passive cloud \citep[e.g.,][]{2004MNRAS.351.1151B,2006MNRAS.373..469B,Peng:2010gn,2014ApJS..214...15S,2015Renzini}. While the overall distribution is consistent with previous findings, a key difference is the absence of the low-mass peak in the red (quenched) population. In \citet{2015Renzini}, the quenched galaxy population exhibits a double-peaked mass distribution, with the low-mass peak attributed to environmental quenching processes affecting satellite galaxies. However, this study focuses exclusively on central galaxies, where environmental effects are less significant. As a result, the low-mass peak of the red cloud is missing, reflecting the intrinsic differences between the central and satellite galaxy populations.

The right panel of Figure A1 shows the distribution of these galaxies on the R$_{\rm disk}$–M$_*$ plane. Both red and blue central spiral galaxies lie along a single mass-size relation. This contrasts with previous findings where quenched galaxies as a whole exhibit systematically smaller sizes than star-forming galaxies of the same stellar mass, with red and blue galaxies following distinct mass-size relations \citep[e.g.,][]{2014ApJ...788...28V,2016ARA&A..54..597C,2019ApJ...880...57M}. The discrepancy arises because, in the high-mass regime, passive galaxies are predominantly more compact due to the significant contribution of elliptical galaxies to the quenched population. However, our sample is restricted to spiral galaxies, which inherently have larger sizes than ellipticals. Consequently, the observed single sequence for red and blue central spirals reflects the specific morphological focus of this study.

From a physical perspective, the choice to focus on central spiral galaxies was deliberate to minimize external environmental effects, including those arising from mergers or other interactions. Processes such as feedback-driven quenching or halo-quenching, which are not expected to disrupt the morphology of galaxies significantly, were of particular interest. The alignment of red and blue central spirals along a single mass-size relation supports the hypothesis that, in the absence of mergers, quenching in central spirals may be governed by mechanisms such as feedback or halo-quenching. These processes allow the disk structure of galaxies to remain largely intact, preserving their spiral morphology regardless of star formation activity. This finding underscores the importance of isolating central spirals to better understand the intrinsic quenching mechanisms at play.

In summary, Figure A1 reveals that central spiral galaxies exhibit a bimodal distribution on the SFR-M$_*$ plane, consistent with the general galaxy population. However, they follow a unified mass-size relation regardless of their star formation activity. The preservation of disk sizes among both passive and star-forming central spirals indicates that quenching mechanisms in these galaxies operate without significant morphological transformation.


\end{document}